\documentclass[12pt]{iopart}
\usepackage{iopams}

\expandafter\let\csname equation*\endcsname\relax

\expandafter\let\csname endequation*\endcsname\relax

\usepackage{graphicx, hyperref, geometry, footmisc, amsmath, mdframed, revsymb}
\usepackage{calligra}
\usepackage{caption}
\usepackage{subcaption}
\DeclareMathAlphabet{\mathcalligra}{T1}{calligra}{m}{n}
\DeclareFontShape{T1}{calligra}{m}{n}{<->s*[2.2]callig15}{}
\newcommand{\scriptr}{\mathcalligra{r}\,}

\newcommand{\scriptt}{\mathcalligra{t}\,}

\begin{document}

\title[]{An exact approach to frustrated total internal reflection and Goos-H$\ddot{\rm\bf a}$nchen shift for $s$-polarised light}

\author[1]{Sayanho Biswas}
\address{IISER Kolkata}
\ead{sb21ms164@iiserkol.ac.in}
%{saibabaxaverian@gmail.com}

\author[2]{Kolahal Bhattacharya
\footnote{Author to whom any correspondence should be addressed: kolahalbhattacharya@sxccal.edu}
}

\address{St Xavier's College Kolkata}
\ead{kolahalbhattacharya@sxccal.edu}

\vspace{10pt}
\begin{indented}
\item[]
\end{indented}

\begin{abstract}
An exact analogy between wave mechanics in quantum theory and the scalar wave treatment of optics emerges from the marriage of Newtonian formulation of geometrical optics~\cite{10.1119/1.14861} and the ``formal quantum theory of light rays''~\cite{Gloge:69}. Here the incidence of a ray of light on the interface between two media is treated as the incidence of a wavefunction on a potential barrier. This leads to the coefficient of reflection identical to Fresnel's formula for $s$-polarised light~\cite{EJP2022}, although there is no concept of the polarisation of light in this model. In the present work, we apply this model to the total internal reflection of light and evanescent waves. We also deduce a one-to-one correspondence between the transmission coefficient in wave mechanics and frustrated total internal reflection for all angles of incidence. Further, we demonstrate that this model can also be used to derive the Goos-H$\ddot{\rm a} $nchen shift for $s$-polarised light. This work augments the discussion on these topics found in the standard texts in optics~\cite{hecht2002optics}.
\end{abstract}
%Though there is no concept of polarisation of light in this model, it leads to the coefficient of reflection identical to Fresnel's formula for $s$-polarised light~\cite{EJP2022}, when the incidence of a ray of light on the interface between two media is treated as the incidence of a wavefunction on a potential barrier. 

%\maketitle
%\hrule
%\pagenumbering{arabic}
%\tableofcontents
\vspace{12pt}
%\hrule

%https://iopscience.iop.org/article/10.1088/0143-0807/29/6/N02

%https://iopscience.iop.org/article/10.1088/1361-6404/ad4b76

\section{Introduction}
% Here our objective is to use the quantum analogy for the scalar wave theory of light to show its implications in two prime examples, namely, the evanescent wave, and the phenomenon of frustrated total internal reflection (FTIR). We shall use the analogy developed to solve equivalent 1-D quantum mechanical problems and derive well known equations in wave optics without having apriori knowledge of Maxwell's equations.

The scalar wave treatment of light emerges during the transition from geometrical optics to physical optics. This model finds applications in the quantitative description of interference and diffraction of light waves due to conceptual simplicity. However, one cannot describe the polarisation of light using this model. This is because the scalar wave treatment emerges from the ray picture of light which lacks the concept of polarisation. Gloge and Marcuse~\cite{Gloge:69} developed the scalar wave treatment starting from Fermat's principle and gave it the form of a quantum theory of light rays. The wavefunction of this theory satisfies the reduced wave equation whose solutions are plane and spherical waves. 

Seventeen years later, a very interesting work on geometrical optics was put forward by Evans and Rosenquist~\cite{10.1119/1.14861}. Starting from Fermat's principle, they showed that one can formulate geometrical optics as an initial value problem like Newton's second law of motion F=ma, where symbols carry their usual meaning. In that paper, they also gave a table of correspondences between the variables commonly used in mechanics and geometrical optics. For example, the optical equivalents of mass, velocity, potential energy and total energy are unity, refractive index ($n$), $-n^2/2$ and zero respectively. It is important to understand that the dimensions of optical equivalent quantities are separate from the corresponding quantities in mechanics. 

Though each of these works was cited extensively in the literature, the connection between them was perhaps not recognized for a long time. Very recently, it has been realized~\cite{EJP2022} that it is possible to treat the reduced wave equation (deduced by Gloge and Marcuse) as a Schr$\ddot{\rm o}$dinger equation where the light ray passing through a medium of refractive index $n$, can be represented by a wavefunction moving through a potential $-n^2/2$ with zero eigenvalue. Gloge et al. used their model to describe the paraxial rays, by treating the $z$ coordinate as a pseudo-time variable and approximating the derivatives of $x$ and $y$ coordinates with respect to $z$ to be small. Their model was called the formal quantum theory of light rays. On the contrary, the model described in~\cite{EJP2022} can perhaps be called the exact or unapproximated quantum theory of light rays, since there is no approximation here. It was shown that the unapproximated model can be used to prove Snell's law in geometrical optics. Using standard quantum mechanical techniques, one can also use this model to estimate the coefficient of reflection when a ray of light is incident on an interface between two media with different refractive indices. Even though the polarisation of light is not included in this model, the formula for the reflection coefficient predicted by this model matches exactly with the corresponding Fresnel's equation for $s$-polarised light. It is not surprising that this model is unable to predict the corresponding formula for the $p$-polarised light. After all, the ray picture does not have the full information on the polarisation of light.

This may appear to be a happy coincidence in a simple case. However, in this paper, we would like to argue that this observation is not limited to results at the elementary level. Rather this unapproximated model can be used further to describe topics like the evanescent waves and frustrated total internal reflection (FTIR). The former was discovered in 1965~\cite{hirschfeld1965total}, but it drew pedagogical interests in recent times~\cite{ morales2024properties, lakhtakia2008total}. The latter was known since Newton's time but was recorded in modern literature form in the late nineteenth century by Sir J C Bose~\cite{bose1898influence}. In FTIR, we find energy flow parallel to the interface in the second medium, but there is no energy flow perpendicular to the interface. After a century, this subject culminated in an active research topic in various scientific fields~\cite{taitt2005evanescent,bliokh2014extraordinary,shirota2017measuring} and created interests in physics pedagogy~\cite{zhu}. We shall show that the unapproximated quantum theory of light rays can be easily applied to the total internal reflection to deduce the form of the evanescent wave and the transmission coefficient for $s$-polarised light in the case of FTIR. It is important to note that previous authors who applied standard quantum mechanics to address these topics, also observed that the results were compatible with $s$-polarised light. For example, Zhu~\cite{zhu} showed a correspondence between the coefficient of transmission in the case of FTIR with $s$-polarised light and the case of transmission of a wavefunction through a one-dimensional potential barrier for an angle of incidence $\theta_0=45^o$. However, in his words, ``the correspondence is not exact for $p$-polarisation''. In this case, however, the energy of the incoming wavefunction is assumed to be non-zero which is inconsistent with the tenets of the unapproximated quantum theory of light rays. In this paper, we will also clarify the cause of this apparent inconsistency.

A lateral shift occurs during the FTIR which was first observed in 1947 by Goos and H$\ddot{\rm a}$nchen~\cite{goos1947new}, called the Goos-H$\ddot{\rm a}$nchen shift. Both FTIR as well as Goos-H$\ddot{\rm a}$nchen shift find applications in fibre optics~\cite{addanki2018review, mackanos2010fiber}. Specifically, Goos-H$\ddot{\rm a}$nchen shift has promising applications in optical sensors and nano-photonics. Standard texts in optics (see chapter 4 and chapter 5 of the book by Hecht~\cite{hecht2002optics}, and page 52 of Born and Wolf~\cite{born2013principles} etc.) mention this shift, but the mathematical details are not discussed. Artmann~\cite{artmann1948berechnung} deduced the formulae for this shift for light with $s$ and $p$-polarisations using principles of physical optics. On the other hand, the quantum mechanical treatment of the Goos-H$\ddot{\rm{a}}$nchen shift by Hora~\cite{hora1960side} leads to the expression which is the same as the result found by Artmann for the $s$-polarised light. In this work, we shall use the unapproximated quantum theory of light rays and deduce the Goos-H$\ddot {\rm a}$nchen shift for $s$-polarised light from a complex reflection amplitude of FTIR. We hope this exercise will convince the readers about the validity and reach of this model.

%There was a little confusion about the validity of Artmann's results in the past~\cite{renard1964total,lotsch1968reflection,carniglia1976correction}, but currently his results are accepted~\cite{lai1986goos, shi2007theory}. Such complex reflection amplitude has applications in thin film analysis and the design of metamaterials. 
We begin our discussion by showing how this unapproximated model predicts the form of evanescent waves in the case of total internal reflection in the following section~\ref{S2}. In the following section~\ref{S3}, we first discuss how the coefficient of transmission for FTIR can be determined using the model. Then we point out the similarities to and differences from the existing work by Zhu~\cite{zhu}. In the final section~\ref{S4}, we apply the model to deduce the Goos-H$\ddot{\rm a}$nchen shift and note the compatibility with Hora's results~\cite{hora1960side}. We conclude with some general discussion on the model and its relevance in pedagogy of physics in the last section~\ref{S5}.

% In 1964, Renard~\cite{renard1964total} pointed out that these formulae predicted a non-zero shift for grazing incidence, which was contrary to expectation. A few years later, Lotsch~\cite{lotsch1968reflection} was able to derive formulae for the shifts that vanished for grazing incidence. Within a decade, however, Carniglia~\cite{carniglia1976correction} showed that a corrected version of Lotsch's work is consistent with Artmann's results. Further scholarly works by Lai~\cite{lai1986goos} and Shi~\cite{shi2007theory} supported the results obtained by Artmann.

\section{Evanescent Waves}\label{S2}
% We directly start off with the equations derived for reflection and refraction in a paper by Prof. Bhattacharya\cite{EJP2022}. The translation invariance of the problem (discussed in the paper) in the y-direction demands

Let us consider the reflection of a light ray from an interface between two media of different refractive indices $n_0$ and $n_1$, as shown in the following Fig.~\ref{fig1a}. On the right-hand side Fig.~\ref{fig1b}, we show the corresponding mechanically equivalent diagram. This figure should be interpreted as the following. A wavefunction representing a ray of light is moving through a potential $V_0=-n_0^2/2$. It is incident at the boundary of another potential $V_1=-n_1^2/2$. Hence, an optical equivalent of force operates on the ray represented by the wavefunction in the $x$ direction. The optical equivalent of the momenta of the ray in the two media are respectively given by $k_0= \sqrt{2m(E-V_0)} /\hbar$ and $k_1=\sqrt{2m(E-V_1)}/\hbar$. Since there is no force in the $y$ direction, the $y$ component of the momentum stays the same: 

$$k_{0_{y}} = k_{1_{y}} \Rightarrow k_0 \sin\theta_0 = k_1 \sin\theta_1$$
$$\Rightarrow (\sqrt{2m(E - V_0)}/\hbar) \sin\theta_0 = (\sqrt{2m(E - V_1)}/\hbar) \sin\theta_1$$

\noindent Substituting $m=1$, $E = 0$ and $V_i = -\frac{n_i^2}{2}$ in accordance with the unapproximated quantum theory of light rays~\cite{EJP2022}, we obtain Snell's law:

\begin{equation}\label{Eq1}
    \boxed{n_0\sin\theta_0 = n_1\sin\theta_1}
\end{equation}
%%%%%%%%%%%%%%%%%%%%%%%%%%%%%%%%%%%%%%%%%%%%%
\begin{figure}[h]
    \centering
    \begin{subfigure}{0.5\textwidth}
    \centering
        \includegraphics[width=1.0\linewidth]{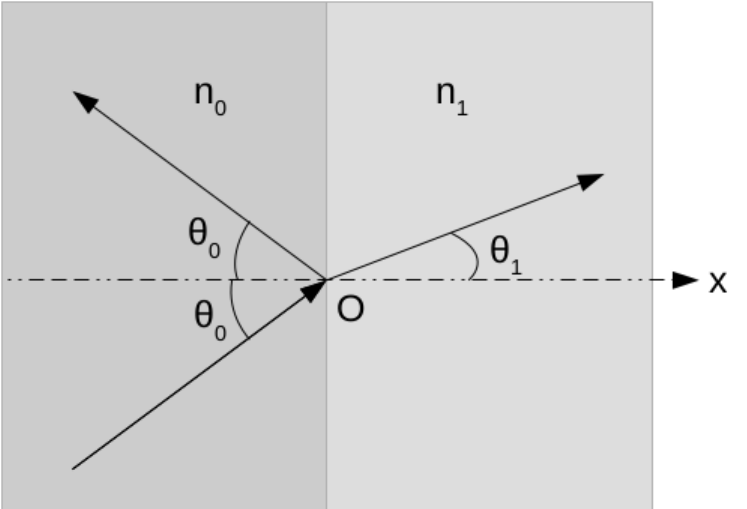}
        \caption{}
        \label{fig1a}
    \end{subfigure}%
    \begin{subfigure}{0.475\textwidth}
    \centering
        \includegraphics[width=1.0\linewidth]{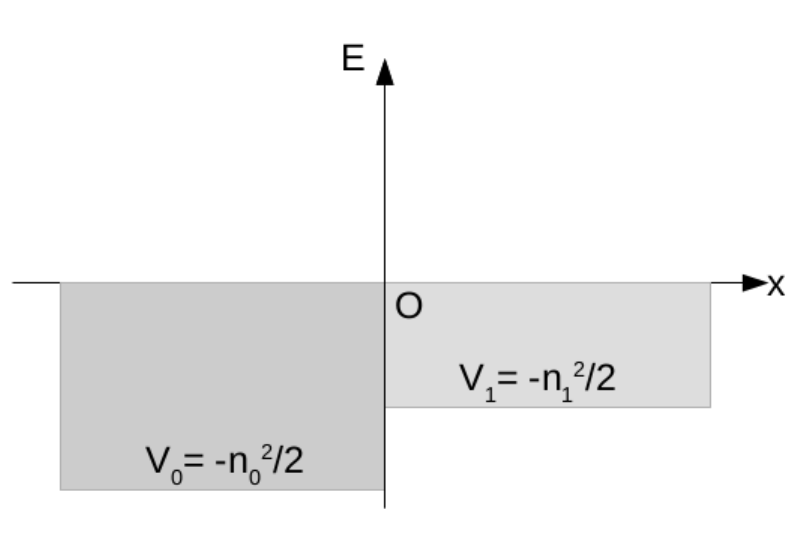}
        \caption{}
        \label{fig1b}
    \end{subfigure}
\caption{(a) Light ray incident on an interface between two media of refractive indices $n_0$ and $n_1$; (b) The unapproximated quantum mechanical description of the problem. Light ray has an optical equivalent of total energy zero; the piece-wise constant potential in the two media are $V_0=-n_0^2/2$ and $V_1=-n_1^2/2$.}
\label{fig1}
\end{figure}
\\
Now, we shall look at the $x$ component of the wave in the medium with refractive index $n_ 1$ and the corresponding potential $V_1$
\begin{equation}\label{Eq2}
    \begin{split}
        k_{1_x} &= k_1 \cos\theta_1 = \pm k_1 \sqrt{1 - \sin^2 \theta_1}\\
        & = \pm k_1 \sqrt{1 - \frac{\sin^2 \theta_0}{(n_1/n_0)^2}}
    \end{split}
\end{equation}
Let us assume $n_0 > n_1$. Then, for the critical angle of incidence ($\theta_c$),

$$\sin\theta_c = \frac{n_1}{n_0}$$
For total internal reflection, we must have

$$\theta_0 > \theta_c \Rightarrow \sin\theta_0 > \sin\theta_c$$
$$\Rightarrow \sin\theta_0 > \frac{n_1}{n_0}$$
Using this fact we can rewrite Eq.\eqref{Eq2} as
\begin{align}\label{Eq3}
    k_{1_x} &= \pm i k_1 \sqrt{\frac{\sin^2 \theta_0}{(n_1/n_0)^2} - 1}\nonumber\\
    \Rightarrow\quad k_{1_x} &= \pm i\frac{k_1}{n_1}\alpha,
\end{align}
where $\alpha = \sqrt{n_0^2\sin^2\theta_0 - n_1^2}$. Thus the solution to the transmitted wave is
\begin{align*}
    \psi_\mathcal{T} &= t e^{i(\Vec{k_1}\cdot\Vec{r} - \omega t)}\\
    &= t e^{i(k_{1_x}x + k_{1_y}y) - \omega t}\\
    &= t e^{i(\pm i\frac{k_1}{n_1}\alpha x + k_1\sin\theta_1y- \omega t)}\\
    &= t e^{i(\pm i\frac{k_1}{n_1}\alpha x + \frac{k_1}{n_1}n_0\sin\theta_0y- \omega t)}
\end{align*}

\noindent From physical consideration, we cannot have an exponentially diverging solution. So, we take $k_{1_x} = + i\frac{k_1}{n_1}\alpha$. This gives us

\begin{equation}\label{Eq4}
  \boxed{\psi_\mathcal{T} = t e^{-\frac{k_1}{n_1}\alpha x}  e^{i(\frac{k_1}{n_1}n_0\sin\theta_0y- \omega t)}}  
\end{equation}
\noindent Hence, we see the existence of a real, exponentially decaying wave solution in the x-direction. These waves are called evanescent waves.

\section{Application of the model to FTIR}\label{S3}
In this section, we shall first review the coefficient of transmission for the $s$-polarised light in a setup demonstrating frustrated total internal reflection. Then we derive the quantity using the analogous problem of the one-dimensional quantum potential barrier. We shall see that using the unapproximated quantum theory, it is possible to show a one-to-one correspondence between the two cases for all angles of incidence.

\subsection{Transmission Coefficient in Optics}
%%%%%%%%%%%%%%%%%%%%%%%%%%%%%%%%%%%%%%%%%%%%%
\begin{figure}[h]
    \centering
    \begin{subfigure}{0.5\textwidth}
    \centering
        \includegraphics[width=1.0\linewidth]{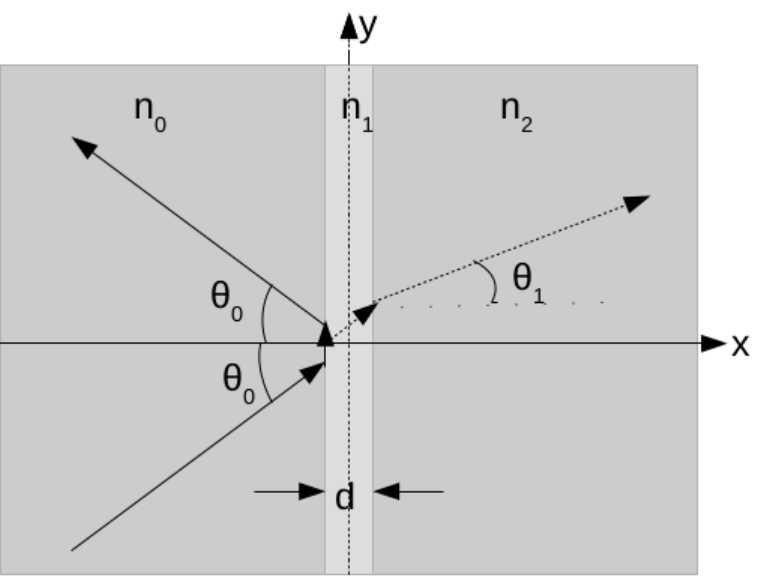}
        \caption{}
        \label{fig1a}
    \end{subfigure}%
    \begin{subfigure}{0.475\textwidth}
    \centering
        \includegraphics[width=1.0\linewidth]{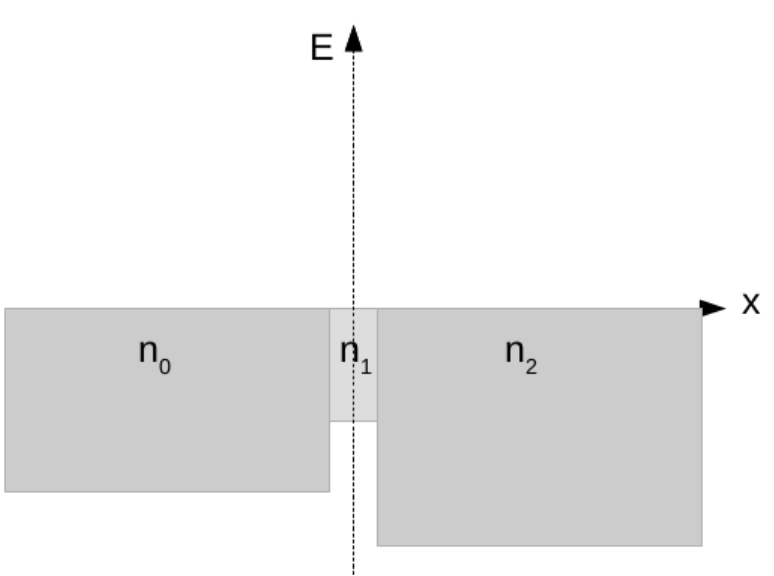}
        \caption{}
        \label{fig1b}
    \end{subfigure}
\caption{(a) A thin medium of thickness $d$ of refractive index $n_1$ sandwiched between media with refractive indices $n_0$ and $n_2$; There is a small lateral shift of the light ray along the interface of the two media, called the Goos-H$ \ddot{\rm a}$nchen shift; (b) The corresponding problem as a wavefunction passing through piece-wise constant one-dimensional potentials $V_0$, $V_1$ and $V_2$.}
\label{fig1}
\end{figure}

\noindent From the paper by I.N Court and F.K von Willisen\cite{Court:64}, we note that the transmission coefficient for a thin film of refractive index $n_1$ in between two semi-infinite dielectric media of refractive indices $n_0$ and $n_2$ can be written as:

\begin{equation}\label{Eq5}
    \mathcal{T} = \frac{1}{\alpha \sinh^2 \xi + \beta}
\end{equation}
where, 
$$\xi = \frac{2\pi d}{\lambda} \sqrt{n_0^2 \sin^2 \theta_0 - n_1^2}$$ 
and for s-polarised light,\\
$$\alpha_s = \frac{(N^2 - 1)(n^2N^2 - 1)}{4N^2\cos\phi_1(N^2\sin^2\theta_0 - 1)\sqrt{n^2 - \sin^2\theta_0}}$$
and 
$$\beta_s = \frac{[\sqrt{n^2 -\sin^2\theta_0} + \cos\theta_0]^2}{4\cos\theta_0 \sqrt{n^2 - \sin^2\theta_0}}$$\\
\noindent Here we have considered $N \equiv n_0/n_1$ and $n \equiv n_2/n_0$. For simplicity, let us assume that $n_2 = n_0$ (i.e. the two semi-infinite dielectric media are made up of the same material). Then we will have $n = 1$. We note that this is taken to just simplify the calculations, but the results will hold for any general case of $n_1 \neq n_2 \neq n_3$. Now we shall substitute the values of $\alpha_s$ and $\beta_s$ in Eq.\eqref{Eq5} with the simplification outlined above:

\begin{equation}\label{Eq6}
    \boxed{\mathcal{T} = 1/\left[ 1 + \frac{(n_0^2 - n_1^2)^2 \sinh^2(\frac{d}{\bar\lambda}\sqrt{n_0^2\sin^2\theta_0 - n_1^2})}{4n_0^2\cos^2\theta_0 (n_0^2\sin^2\theta_0 - n_1^2)}\right]}
\end{equation}\vspace{12pt}
-where $\bar\lambda=\lambda/2\pi$.
\subsection{Transmission Coefficient in 1-D Potential Barrier}
We shall now calculate the transmission coefficient for the quantum problem of one-dimensionl square potential barrier. Let us define our potential as follows:

\[ V(x) = \begin{cases} 
      V_0 & x < 0 \\
      V_1 & 0\leq x < d \\
      V_0 & x \geq d
   \end{cases}
\]

\noindent From the Schr$\ddot{\rm o}$dinger equation, we can straightaway write the solution of the time-independent part in the three regions.

\[
    \psi(x) = 
    \begin{cases}
        \psi_1 = A_r e^{ik_0x} + A_l e^{-ik_0x} & x < 0 \\
       %\psi_2 = B_r e^{k_1x} + B_l e^{-k_1x} & 0\leq x < d \\
        \psi_2 = B_r e^{ik_1x} + B_l e^{-ik_1x} & 0\leq x < d\\%KB
        \psi_3 = C_r e^{ik_0x} + C_l e^{-ik_0x} & x \geq d \\
    \end{cases}
\]

\noindent where $k_0 = \sqrt{\frac{2m(E - V_0)}{\hbar^2}}$, $k_1 = \sqrt{\frac{2m(E - V_1)}{\hbar^2}}$.\\From the continuity of the wave function and its first derivative at $x = 0$ and at $x = d$, we have
\begin{subequations}
    \begin{align}
    A_r + A_l &= B_r + B_l\label{Eq7a}\\
    ik_0(A_r - A_l) &= ik_1(B_r - B_l)\label{Eq7b}\\
    B_r e^{idk_1} + B_l e^{-idk_1} &= C_r e^{idk_0} + C_l e^{-idk_0}\label{Eq7c}\\
    ik_1(B_r e^{idk_1} - B_l e^{-idk_1}) &= ik_0(C_r e^{idk_0} - C_l e^{-idk_0})\label{Eq7d}
\end{align}
\end{subequations}
\noindent Now, in order to find the transmission and reflection coefficients, we put $A_r = 1$ (normalized incident particle), $A_l = r$ (reflection amplitude), $C_r = t$ (transmission amplitude) and $C_l = 0$ (no incoming particle coming from the right in region 3). We use Eq.\eqref{Eq7a}--Eq.\eqref{Eq7d} to solve for $t$ and $r$. The transmission amplitude is thus obtained as follows:

\begin{equation}
    t = \frac{4k_0k_1e^{-id(k_0 - k_1)}}{(k_0+k_1)^2 - e^{2idk_1}(k_0-k_1)^2}
\end{equation}

\noindent The transmission coefficient will then be given by:
\begin{align*}
    |t|^2 &= |4k_0k_1|^2 \frac{e^{-id(k_0 - k_1)}}{[(k_0+k_1)^2 - e^{2idk_1}(k_0-k_1)^2]}\frac{e^{id(k_0 - k_1)}}{[(k_0+k_1)^2 - e^{-2idk_1}(k_0-k_1)^2]}\\
    &= \frac{|4k_0k_1|^2}{16k_0^2k_1^2 + 4(k_0^2 - k_1^2)^2\sin^2(k_1d)}\\\\
    \mbox{i.e,}\qquad \mathcal{T} &= 1/\left[ 1 + \frac{4(k_0^2 - k_1^2)^2\sin^2(k_1d)}{(4k_0k_1)^2}\right]
\end{align*}

\noindent To make an analogy with the case of FTIR, which is 2 dimensional in nature, we choose to interpret the above result as being in the $x$-direction. This is a logical assumption since the variation in the refractive index profile is only along $x$-direction. Hence, we shall replace $k_0$ with $k_{0_x} = k_0\cos\theta_0$ and $k_1$ with $k_{1_x} = k_1\cos\theta_1$

$$\mathcal{T} = 1/\left[ 1 + \frac{4(k_{0_x}^2 - k_{1_x}^2)^2\sin^2(k_{1_x}d)}{(4k_{0_x}k_{1_x})^2}\right]$$

$$ = 1/\left[ 1 + \frac{(\frac{2m(E - V_0)}{\hbar^2}\cos^2\theta_0 - \frac{2m(E - V_1)}{\hbar^2}\cos^2\theta_1)^2\sin^2(\sqrt{\frac{2m(E - V_1)}{\hbar^2}}\cos\theta_1d)}{4\frac{2m(E - V_0)}{\hbar^2}\cos^2\theta_0\frac{2m(E - V_1)}{\hbar^2}\cos^2\theta_1}\right]$$\\
From the unapproximated quantum theory (outlined in~\cite{EJP2022}), we make the following replacements:

\begin{table}[h!]
    \centering
    \begin{tabular}{ccc}
         $m$& $\longrightarrow$ & $1$ \\
         $E$& $\longrightarrow$ & $0$ \\
         $V$& $\longrightarrow$ & $-n^2/2$ \\
         $\hbar$& $\longrightarrow$ & $\bar\lambda$ \\
    \end{tabular}
\end{table}

\noindent We then obtain the following expression for the transmission coefficient:

\begin{equation}\label{Eq9}
    \mathcal{T} = 1/\left[ 1 + \frac{(n_0^2\cos^2\theta_0 - n_1^2\cos^2\theta_1)^2 \sin^2(\frac{d}{\bar\lambda}n_1\cos\theta_1)}{4n_0^2\cos^2\theta_0n_1^2\cos^2\theta_1} \right]
\end{equation}
Now, from Eq.\eqref{Eq1},
$$n_0\sin\theta_0 = n_1\sin\theta_1$$
$$\Rightarrow n_0^2(1 - \cos^2\theta_0) = n_1^2(1 - \cos^2\theta_1)$$
Therefore we have
\begin{equation}\label{Eq10}
    n_0^2\cos^2\theta_0 - n_1^2\cos^2\theta_1 = n_0^2 - n_1^2
\end{equation}
and
\begin{equation}\label{Eq11}
    n_1^2\cos^2\theta_1 = n_1^2 - n_0^2\sin^2\theta_0
\end{equation}

\noindent Substituting Eq.\eqref{Eq10} and Eq.\eqref{Eq11} into Eq.\eqref{Eq9}, we find

$$\mathcal{T} = 1/\left[ 1 + \frac{(n_0^2 - n_1^2)^2 \sin^2(\frac{d}{\bar\lambda}\sqrt{n_1^2 - n_0^2\sin^2\theta_0})}{4n_0^2\cos^2\theta_0 (n_1^2 - n_0^2\sin^2\theta_0)}\right]$$

\noindent Here we again make use of the fact that in case $n_0 > n_1$ and $\theta_0 > \theta_c$ (condition for total internal reflection) then we must have $n_0\sin\theta_0 > n_1$. Further we take advantage of the fact that $\sinh (ix) = i\sin x$ to rewrite the above equation

$$\mathcal{T} = 1/\left[ 1 + \frac{(n_0^2 - n_1^2)^2 \sinh^2(\frac{d}{\bar\lambda}\sqrt{n_0^2\sin^2\theta_0 - n_1^2})}{4n_0^2\cos^2\theta_0 (n_0^2\sin^2\theta_0 - n_1^2)}\right]$$

\noindent which is identical to E.\eqref{Eq6}.

\subsection{Earlier Work}
It is worthwhile to note that such a similarity in the transmission coefficient was hinted at in a paper by S. Zhu et al.~\cite{zhu}. However, they were unable to derive a one-to-one correspondence between the two cases for a general angle of incidence. The correspondence made in their paper was, for $\theta_0 = 45^\circ$, $n_1 = 1$ we have:

\begin{align*}
    \frac{mE}{h^2}&\longrightarrow\frac{n_0^2}{4\lambda^2}\\
    \frac{mV_0}{h^2}&\longrightarrow\frac{n_0^2 - 1}{2\lambda^2}
\end{align*}

\noindent which seemed equivalent to the following (for any general angle of incidence $\theta_0$):

\begin{align*}
    &m &&\longrightarrow&& 1 \\
    &\hbar &&\longrightarrow&& \bar\lambda \\
    &V_0 &&\longrightarrow&& -\frac{n_0^2}{2} - \left(-\frac{1}{2}\right) \\
    &E &&\longrightarrow&& \frac{n_0^2}{2}\cos^2\theta_0
\end{align*}

\noindent Although it seems reasonable to consider $V_0$ as the barrier height compared to the surrounding, indicative of the difference in the refractive index of the semi-infinite refractive medium and the thin film, it does not stand to reason that we shall have a non-zero $E$ value. Indeed, making the above transformation would give the transmission coefficient the same as that in FTIR, the origin of the correspondence does not seem as tangible as the one demonstrated in this paper. The question then is, what is the difference between the two correspondences? The answer lies in the fact that the inherent dimensions of the two problems are different. Whereas the FTIR is a two-dimensional problem, Zhu et al.~\cite{zhu} used the  formula for transmission coefficient (from Schiff's test on quantum mechanics) in one-dimension. It is important to convert to the $x$ or $y$-direction depending upon the geometry of the problem. For example, it is imperative that in our case we take $k_{0_x}$ and $k_{1_x}$ in place of $k_0$ and $k_1$ considering the invariance in $y$-direction. This perspective was missed in the paper by Zhu et al.~\cite{zhu} which led to the incorporation of terms involving angles of incidence in the correspondence, leading to a non-zero $E$ value.

\section{Goos-H$\ddot{\rm\bf{a}}$nchen shift}\label{S4}
\begin{figure}
    \centering
\includegraphics[width=0.5\linewidth]{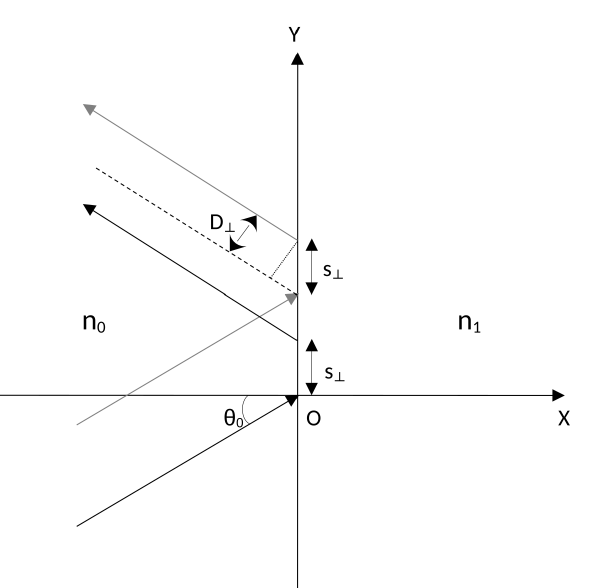}
    \caption{A beam of monochromatic light of wavelength $\lambda$ bounded by black and gray borders is  incident on the interface between two media with refractive indices $n_0$ and $n_1\ (<n_0)$ at an angle $\theta_0$ greater than the critical angle between the two media. The beam is totally reflected into the same media with a lateral shift $s_\perp$ along the $Y$ direction. The Goos-H$\ddot{\rm{a}}$nchen shift is denoted by $D_\perp$. The subscript $_\perp$ corresponds to the $s$ polarised light.}
    \label{Fig3}
\end{figure}

During total internal reflection, a lateral shift of the light beam occurs along the interface of the two media. See Fig.\ref{Fig3} in the following. We attempt to estimate this shift using the unapproximated quantum theory of light rays.
\footnote{Some authors e.g. Hecht~\cite{hecht2002optics,snyder1976goos} identify the shift along the interface as the Goos-H$\ddot{\rm {a}}$nchen shift. But others~\cite{zhu,goos1947new} call the shift in the direction perpendicular to the direction of propagation. We adhere to the latter nomenclature and comment that these are related.}
% , following the work of Hora (which we could not locate) as has been discussed in the \href{http://www.scholarpedia.org/article/Goos-H%C3%A4nchen_effect}{scholarpedia page}. 

We consider an incoming wave packet $\psi_i$ representing a light ray with an optical equivalent of energy $E=0$. It passes through a medium of refractive index $n_0$ in $x<0$ corresponding to an optical equivalent of potential $V_0=-n_0^2/2$. This wavepacket is given by: $\psi_i(x,y)=e^{ik_{0_x}x+ik_ {0_y}y}+\scriptr e^{-ik_{0_x}x+ik_{0_y}y}$. For this wavepacket, we have
\begin{align}
     k_{0_x}&=k_0\cos\theta_0\nonumber\\
     k_{0_y}&=k_0\sin\theta_0\nonumber\\
     k_0\ &=\sqrt{2m(E-{V_0})}/\hbar
\end{align}
We have seen that the transmitted wave has an evanescent component given by: $\psi_t= \scriptt e^{i{k_1}_x x+i{k_1}_y y}=
\scriptt e^{-k_1(\alpha/n_1)x+ik_{1_y}y}$ for $x>0$. For this transmitted wavepacket,
    \begin{align}
     k_{1_x}&=k_1\cos\theta_1\nonumber\\
     k_{1_y}&=k_1\sin\theta_1\nonumber\\
k_1\ &=\sqrt{2m(E-{V_1})}/\hbar\nonumber
\end{align}
Now, let us match the boundary condition at the interface:
\begin{subequations}
\begin{align}
    \psi_i(0,y)&=\psi_t(0,y)\label{eq14a}\\
    \left(\frac{\partial\psi_i}{\partial x} \right)_{0,y}&=\left(\frac{\partial\psi_t}{\partial x}\right)_{0,y}\label{eq14b}
\end{align}
\end{subequations}
Using the forms of the wavefunctions $\psi_i$ and $\psi_t$ in  Eq.\eqref{eq14a}, we see that:
\begin{align}\label{eq15}
    (1+\scriptr)e^{i{k_0}_y y}=\scriptt e^{i{k_1}_y y}\implies 1+\scriptr=\scriptt
\end{align}
On the other hand, Eq.\eqref{eq14b} leads to:
\begin{align}\label{eq16}
    i{k_0}_xe^{i{k_0}_y y}+\scriptr(-i{k_0}_x)e^{i{k_0}_y y}&=\scriptt\left(-k_1\frac{\alpha}{n_1}\right)e^{i{k_1}_y y}\nonumber\\
\implies i{k_0}_x(1-\scriptr)&=-\scriptt k_1 \frac{\sqrt{n_0^2\sin^2\theta_0-n_1^2}}{n_1}
\end{align}
At this point, we invoke the model of `F=ma' optics, and use the observation that momenta $k_0$ and $k_1$ can be expressed as:
\begin{align}
k_0\rightarrow\sqrt{2\cdot1\cdot(0+n_0^2/2)}/\bar\lambda=n_0/\bar\lambda\nonumber\\
k_1\rightarrow\sqrt{2\cdot1\cdot(0+n_1^2/2)}/\bar\lambda=n_1/\bar\lambda
\end{align}
Therefore, from Eq.\eqref{eq16} we find that 
\begin{align}\label{eq18}
\frac{n_0}{\bar\lambda}\cos\theta_0(1-\scriptr)&=i\scriptt\ \frac{n_1}{\bar\lambda} \frac{\sqrt{n_0^2\sin^2\theta_0-n_1^2}}{n_1}\nonumber\\
\implies 1-\scriptr&=i\scriptt \frac{\sqrt{n_0^2\sin^2 \theta_0-n_1^2}}{n_0\cos\theta_0}
\end{align}
From Eq.\eqref{eq15} and Eq.\eqref{eq18}, we can deduce the reflection amplitude $\scriptr $:
\begin{align}
1-\scriptr&=i\frac{\sqrt{n_0^2\sin^2 \theta_0-n_1^2}}{n_0\cos\theta_0}(1+\scriptr)=i\beta(1+\scriptr)\nonumber\\
\implies \scriptr&=\frac{1-i\beta}{1+i\beta}
\end{align}
where $\beta=(\sqrt{n_0^2\sin^2 \theta_0-n_1^2})/(n_0\cos\theta_0)$. Thus the reflection amplitude comes as a constant. Its real part denotes the amplitude of the reflected wave relative to the incident wave and the imaginary part gives the phase shift of the reflected wave with respect to the incident wave. This phase angle is given by:
\begin{align}
\arg(\scriptr)=\Phi_R&=\arg(1-i\beta)-\arg(1+i\beta)\nonumber\\
&=-2\tan^{-1}\beta=-2\tan^{-1}\left(\frac{\sqrt{n_0^2\sin^2 \theta_0-n_1^2}}{n_0\cos\theta _0}\right)
\end{align}
If the beam grazes along through a distance $s_\perp$ along the interface, then the above phase difference can be equated to the integral of $s_\perp$ over $d{k_0}_y$ (where ${k_0}_y=\frac{2\pi}{\lambda}$), because every small increment in $Y$ direction contributes to the phase. In other words, $s_\perp$ is given by:
\begin{equation}
    s_\perp=-\frac{\partial\Phi_R}{\partial {k_0}_y}
\end{equation}
The negative sign appears because the contribution to the phase is less as $s_\perp$ is greater, i.e. we are further away. Then, the GHS for $s$-polarised light is given by:
\begin{align}
D_\perp=s_\perp\cos\theta_0=\cos\theta_0\left(-\frac{\partial\Phi_R}{\partial {k_0}_y}\right)
\end{align}
The differential of ${k_0}_y$ can be written in terms of $\theta_0$ as: 
\begin{equation}
d{k_0}_y=k_0\cos\theta_0 d\theta_0=(n_0/\bar\lambda)\cos\theta_0 d\theta_0
\end{equation}
Therefore, the said shift is given by:
\begin{align}
    D_\perp=-\frac{\bar\lambda}{n_0}\frac{d\Phi_R}{d\theta_0}
\end{align}
Using the expression of the phase of the amplitude of reflection, we find that:
\begin{equation}
    D_\perp=+\frac{\lambda}{2\pi n_0}\left(2\frac{n_0\sin\theta_0}{\sqrt{n_0^2\sin^2\theta_0-n_1^2}}\right) =\frac{\lambda}{\pi}\frac{\sin\theta_0}{\sqrt{n_0^2\sin^2\theta_0-n_1^2}}
\end{equation}
This expression is the same as the one found by Artmann~\cite{artmann1948berechnung} and later by Hora~\cite{hora1960side} for the $s$-polarised light. 
%%%%%%%%%%%%%%%%%%%%%%%%%%%%%%%%%%%%%%%%%%%%%%%%%%
\section{Usefulness in the pedagogy of physics}\label{S5}
In this paper, we used the unapproximated quantum theory of light rays to address the topics of frustrated total internal reflection (FTIR), evanescent waves and Goos-H$\ddot{\rm a}$nchen shift that occurs during FTIR. The model is based on the pioneering works~\cite{10.1119/1.14861,Gloge:69} on the analogy between optics and mechanics, and was introduced in~\cite{EJP2022}. The major success of the model is to establish a one-to-one connection between the transmission coefficient in FTIR and the unapproximated model in this paper for all the values of the incident angle. Zhu et al.~\cite{zhu} found the correspondence only at $\theta_0=45^o$, perhaps because he did not treat the phenomena in two orthogonal directions separately. But it is a remarkable observation that Zhu et al.~\cite{zhu} and Hora~\cite{hora1960side} both found that the correspondence comes only for $s$-polarised light. We comment that the results predicted based on the unapproximated model are also limited to $s$-polarised light. Whether or not it is possible to conceive a consistent description of polarisation of light from the unapproximated quantum theory of light rays, is an open problem for now. 

This work complements the discussion on total internal reflection and evanescent waves found in textbooks. It should also be useful to the students and teachers taking fibre optics and/or photonics courses. Most importantly, this work makes an important contribution to the list of correspondence between wave optics and quantum mechanics which should draw the attention of the physics community. 

% \section{Acknowledgement} The authors acknowledge the National Initiative for Undergraduate Sciences (NIUS) program conducted by the Homi Bhabha Centre for Science Education (TIFR) for hosting the NIUS camp and supporting the research in the form of associating the student with his mentor and supporting travel, etc.

\section{Data availability statement} 
No new data were created or analysed in this work. All the information pertaining to the work is included in this paper.

\section{Author interests statement}
The authors declare no conflict of interests.

\section{Acknowledgments}
The authors thank the NIUS program (National Initiative for Undergraduate Sciences) conducted by the Homi Bhabha Centre for Science Education (TIFR), Mumbai, for providing an academic platform to carry out this research. The research work was not funded by any funding agency. 

% We have thus shown an important application of the quantum analogy for the scalar wave equation of light in deriving the expression for transmission coefficient in the case of FTIR. However, we note that the applicability of this analogy yet remains within the regime of $s$-polarised light. A consistent description for $p$-polarised electromagnetic waves is yet to be discovered and is an open problem for now.
%\newpage

\section{Bibliography}

\bibliographystyle{unsrt}%plainurl
\bibliography{report}

\end{document}